\documentclass[%
 reprint,
superscriptaddress,
 amsmath,amssymb,
prb,
]{revtex4-2}

\usepackage{graphicx}
\usepackage{dcolumn}
\usepackage{bm}

\begin{document}

\preprint{APS/123-QED}

\title{Determining the bulk and surface electronic structure of $\alpha$-Sn/InSb(001) with spin- and angle-resolved photoemission spectroscopy}

\author{Aaron N. Engel}
\email{aengel@ucsb.edu}
\affiliation{Materials Department, University of California Santa Barbara, Santa Barbara, California 93106, USA}

\author{Paul J. Corbae}
\affiliation{Electrical and Computer Engineering Department, University of California Santa Barbara, Santa Barbara, California 93106, USA}

\author{Hadass S. Inbar}
\affiliation{Materials Department, University of California Santa Barbara, Santa Barbara, California 93106, USA}

\author{Connor P. Dempsey}
\affiliation{%
Electrical and Computer Engineering Department, University of California Santa Barbara, Santa Barbara, California 93106, USA}%

\author{Shinichi Nishihaya}
\altaffiliation[]{Present Address: Department of Physics, Tokyo Institute of Technology, Tokyo 152-8551, Japan}
\affiliation{%
 Electrical and Computer Engineering Department, University of California Santa Barbara, Santa Barbara, California 93106, USA}%

\author{Wilson Y\'{a}nez-Parre\~{n}o}
\affiliation{%
 Electrical and Computer Engineering Department, University of California Santa Barbara, Santa Barbara, California 93106, USA}%

\author{Yuhao Chang}
\affiliation{Materials Department, University of California Santa Barbara, Santa Barbara, California 93106, USA}

\author{Jason T. Dong}
\affiliation{Materials Department, University of California Santa Barbara, Santa Barbara, California 93106, USA}

\author{Alexei V. Fedorov}
\affiliation{
Advanced Light Source, Lawrence Berkeley National Laboratory, Berkeley, California 94720, USA
}%
\author{Makoto Hashimoto}
\affiliation{
 Stanford Synchrotron Radiation Lightsource, SLAC National Accelerator Laboratory, 2575 Sand Hill Road, Menlo Park, California 94025, USA
}%
\author{Donghui Lu}
\affiliation{
 Stanford Synchrotron Radiation Lightsource, SLAC National Accelerator Laboratory, 2575 Sand Hill Road, Menlo Park, California 94025, USA
}%

\author{Christopher J. Palmstr\o{}m}
\email{cjpalm@ucsb.edu}
\affiliation{Materials Department, University of California Santa Barbara, Santa Barbara, California 93106, USA}
\affiliation{%
 Electrical and Computer Engineering Department, University of California Santa Barbara, Santa Barbara, California 93106, USA}%


\begin{abstract}
The surface and bulk states in topological materials have shown promise in many applications. 
Grey or $\alpha$-Sn, the inversion symmetric analogue to HgTe, can exhibit a variety of these phases. 
However there is disagreement in both calculation and experiment over the exact shape of the bulk bands and the number and origin of the surface states. Using spin- and angle-resolved photoemission we investigate the bulk and surface electronic structure of $\alpha$-Sn thin films on InSb(001) grown by molecular beam epitaxy. 
We find that there is no significant warping in the shapes of the bulk bands. We also only observe the presence of two surface states near the valence band maximum in both thin (13 bilayer) and thick (400 bilayer) films. 
In 50 bilayer films, these two surface states coexist with quantum well states. 
Surprisingly, both of these surface states are spin-polarized with orthogonal spin-momentum locking and opposite helicities. 
The attribution of the spin polarization to these two surface states is verified and potential origins are discussed. Finally, the presence of a second orthogonal spin-momentum locked topological surface state deep below the valence band maximum is verified. Our work clarifies the electronic structure of $\alpha$-Sn(001) to allow for better agreement between experiment and calculation such that better control of the electronic properties can be achieved. In addition, the presence of two spin-polarized surface states has important ramifications for the use of $\alpha$-Sn in spintronics. 
\end{abstract}

\maketitle


\section{\label{sec:level1}Introduction}

$\alpha$-Sn, the diamond allotrope of Sn, has long been known to have an inverted band structure \cite{Groves1963}. In contrast to similarly diamond structured group IV semiconductors Si and Ge, in $\alpha$-Sn relativistic effects are strong such that the large 5$p$-splitting from the spin-orbit interaction large band shifts between the 5$s$ and 5$p$ states via the mass Darwin effect results in an inverted band structure \cite{kufner2015}. The band inversion is between the second valence band and the conduction band analogous to HgTe.  In the bulk $\alpha$-Sn is a gapless semiconductor, where the quadratic touching point between the $p$-like $\Gamma_{8,c}^+$ and $\Gamma_{8,v}^+$ is assured by the symmetries of the system.

Bulk $\alpha$-Sn is not stable at room temperature, transitioning to the topologically trivial, superconducting $\beta$-Sn above 13.2\textdegree C \cite{TinPest}. However, it was found that by growing thin films via molecular beam epitaxy on closely lattice- and symmetry-matched substrates, the transition temperature could be raised above room temperature \cite{Farrow1981}. The exact transition temperature varies significantly based on strain and film thickness \cite{Song2019,Fitzgerald1991,Farrow1981,Reno1989}, but is typically greater than 50\textdegree C and can reach up to 200\textdegree C. It is now generally accepted that compressive strain from epitaxial growth of $\alpha$-Sn on the common substrates of InSb(001) (-0.15\%) or CdTe(001) (-0.12\%) results in  the formation of a 3D Dirac semimetal (DSM) phase. The crossing is between the $\Gamma_{8,c}^+$ and $\Gamma_{8,v}^+$ bands in the $[001]$ direction and is enforced by $C_4(z)$ symmetry \cite{Anh2021,Chen2022,deCoster2018,Huang2017}. Tensile strain is expected to result in a 3D topological insulator (TI) phase \cite{Carrasco2018,Zhang2018}, however (to our knowledge) this has not yet been realized experimentally as a suitable substrate has not been demonstrated. These strain-based transitions are shown via a tight-binding model in Section S1 of the Supplemental Material \cite{Supp}.

The effect of quantum confinement on $\alpha$-Sn thin films has also been of much interest. Compressively strained ultrathin films have been suggested to be a 3D TI \cite{Barfuss2013,Ohtsubo2013,Kufner2014} or 2D TI \cite{Anh2021} when grown on InSb(001), a 2D DSM \cite{Xu2017}  or 2D TI  \cite{Xu2018,Rogalev2019} when grown on InSb(111), or a 2D TI phase coexistent with unconventional superconductivity when grown on PbTe(111)/Bi$_2$Te$_3$(0001) \cite{Falson2020}. Many of these studies involve either surface dosing or bulk doping with Te or Bi to improve surface quality and electron-dope the (usually degenerately $p$-type doped) films  \cite{Barfuss2013,Scholz2018,Ohtsubo2013}, which could have unexpected effects on the surface electronic structure and has been consistently measured to modulate at least the band velocity of the topological surface states \cite{Madarevic2020,engel2023growth, Chen2022}. 

The tunability between these phases in $\alpha$-Sn make it an interesting testbed for topological phase transitions. There are a limited number of topological materials which may be grown as high quality thin films, especially at the ultrathin limit where quality usually degrades.  
These, typically compound, materials are prone to point defects and it can be difficult to achieve the desired stoichiometry or chemical potential, which can sometimes be solved via alloying, doping, or electrostatic gating. $\alpha$-Sn, an elemental semimetal/semiconductor, should not suffer from many of these problems. Particularly interesting for spintronic applications, $\alpha$-Sn has shown very efficient spin-charge conversion at room temperature, showing much promise for applications involving current-induced spin-orbit torques and other spintronic-based devices \cite{Sanchez2016,Ding2021a,Ding2021b,Zhang2022spin}.

Although there has been much interest in $\alpha$-Sn, there are still open questions in the band structure which we seek to answer here. In early band structure measurements of $\alpha$-Sn thin films via cyclotron resonances in magnetotransport and magnetotransmission \cite{Hoffman1989} and angle-resolved photoemission spectroscopy \cite{hochst1983,hochst1985,Middelmann1987,Tang1987}, the presence of topological surface states was not observed. In more recent angle-resolved photoemission measurements of $\alpha$-Sn/InSb(001) thin films by Barfuss \textit{et al.} \cite{Barfuss2013}, one topological surface state with the expected helical spin-momentum locked spin polarization of a Dirac-like surface state was observed while doping the films with percent levels of Te. Soon after, Ohtsubo \textit{et al.} \cite{Ohtsubo2013} similarly measured the topological surface state, but with an opposite helicity for the spin texture and using an adlayer of Bi on the $\alpha$-Sn(001) surface. Further work found topological surface states on $\alpha$-Sn/InSb(111) as well, however the spin polarization of these states has not yet been measured \cite{Xu2017,Rogalev2019}. More work on the (001) surface of Te-doped $\alpha$-Sn thin films found the presence of a second, unpredicted surface state \cite{Scholz2018}. Building on this study, Chen \textit{et al.} \cite{Chen2022} found evidence of three surface states in undoped $\alpha$-Sn(001), of which one was attributed to be the typical topological surface state and the other two were associated with a Rashba-split surface state with a large Rashba coefficient. In addition, trivial surface states associated with the surface reconstruction of $\alpha$-Sn(001) have been proposed \cite{Lu1998} of which there is reasonable experimental agreement via surface core level shifts in ultraviolet photoelectron spectroscopy measurements \cite{Cricenti2001}.

Finally, the location of the Dirac node of the “typical” topological surface state in  $\alpha$-Sn(001) is not robustly known. 
The crossing is commonly calculated to be in the middle of the $\Gamma_{8,v}^+$-$\Gamma_{7}^-$ gap, independent of film thickness \cite{Barfuss2013,Chen2022,Jardine2023}, while many experiments find the crossing to be significantly closer to the valence band maximum, if not above the valence band maximum \cite{Rogalev2017,Chen2022,engel2023growth}. 
There are limited direct measurements of the distance between the surface state Dirac node and the valence band maximum at the $\mathbf{\Gamma}$ point. 

Our primary focus in this work is to investigate the number, the nature, and the dispersion of surface states of compressively strained $\alpha$-Sn/InSb(001) via spin- and angle-resolved photoemission spectroscopy (SARPES and ARPES). We first investigate the relationship between the dispersion of the surface states and the bulk bands in ultrathin films where confinement effects are strong. The existence of only two surface states near the valence band maximum is confirmed, both of which terminate in the confinement-induced bulk band gap. Only these two surface states are present across a large range of film thicknesses investigated here. We further measure an additional band inversion and topological surface state deep below the Fermi level, consistent with the work of Rogalev \textit{et al.} \cite{Rogalev2017}. 
Interestingly, \textit{both} of the observed surface states near E$_\mathrm{F}$ are revealed to be spin-polarized with the expected ideal orthogonal spin-momentum locking and opposite helicities.
Our results clarify the surface and bulk electronic structure of $\alpha$-Sn and challenge the results of many calculations. 
The presence of two surface states with opposite polarization has an important bearing on future spin-charge conversion measurements as the position of the surface Fermi level can drastically change the spin polarization of the Fermi surface and thus the spin-charge conversion efficiency.
\begin{figure}[h]
\includegraphics{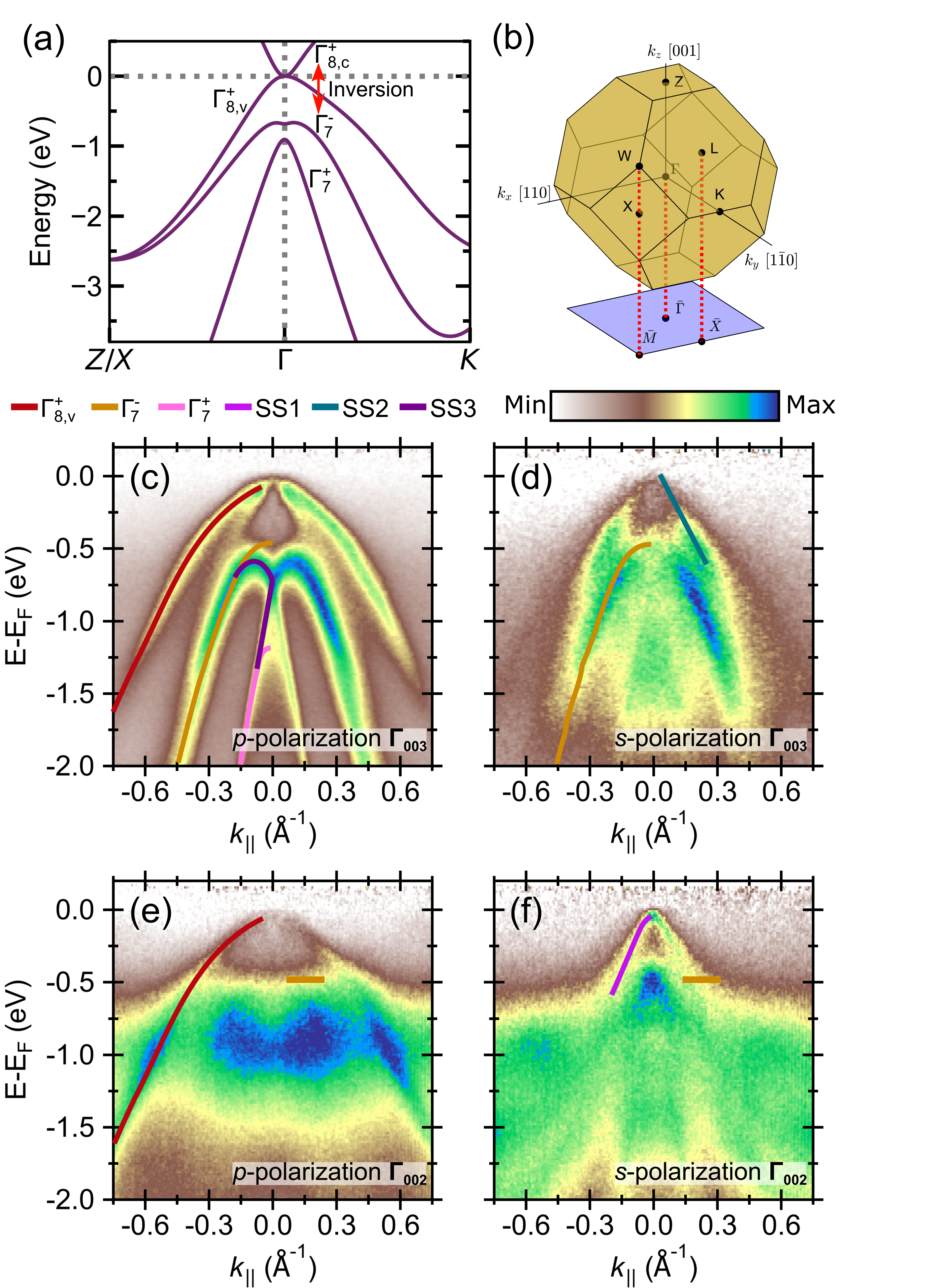}
\caption{\label{Fig1} (a) Electronic structure of $\alpha$-Sn as calculated by the tight-binding model modified from \cite{Rogalev2017}. The band inversion is indicated. (b) Bulk and surface Brillouin zone for the (001) orientation of $\alpha$-Sn. ARPES measurements on 13 BL of $\alpha$-Sn at (c) \bm{$\Gamma_{003}$} with $p$-polarized light (d) \bm{$\Gamma_{003}$} with $s$-polarized light (e) \bm{$\Gamma_{002}$} with $p$-polarized light and (f)  \bm{$\Gamma_{002}$} wth $s$-polarized light.  \bm{$\Gamma_{003}$} corresponds to $h\nu$=127 eV and  \bm{$\Gamma_{002}$} corresponds to $h\nu$=55.8 eV. Guides to the eye for surface states and valence bands are indicated. Horizontal lines correspond to the maximum of the associated band. All measurements are along the $\overline{X}-\overline{\Gamma}-\overline{X}$ direction.}
\end{figure}
\section{Methods}
Thin films of $\alpha$-Sn were grown on both the indium rich c(8$\times$2) and the antimony rich c(4$\times$4) reconstruction of InSb(001) as discussed in our earlier work \cite{engel2023growth}. Variable growth temperature and rates were used, as described previously \cite{engel2023growth}. Film thicknesses are referred to in bilayers (BL) where 1 BL corresponds to half of the conventional diamond cubic unit cell (1 BL = 9.5$\times 10^{14}$ at/cm$^2$). In this work, a single 13 BL film and a single 50 BL film are studied. Four different 400 BL films labelled 400 BL-A,B,C,D are studied as well. The surface reconstruction of $\alpha$-Sn films as measured by reflection high energy electron diffraction generally showed the mixed (2$\times$1)/(1$\times$2) reconstruction, except 400 BL-B which showed a (2$\times$2) reconstruction. Tight-binding calculations were performed in the Chinook framework \cite{Day2019} with parameters slightly modified from those reported in Ref. \cite{Rogalev2017} which were themselves extracted from Ref. \cite{Pollak1970}. Strain was incorporated using Harrison's $d^2$ rule and a modified Harrison's rule for group IV semiconductors \cite{TBstrain}. The tight-binding calculation gives good agreement in Dirac node spacing with experimental results \cite{PolishSn}. Structural characterization of some $\alpha$-Sn films showed high quality films fully strained to the InSb(001) substrate. Further details are given in Section S2 of the Supplemental Material \cite{Supp}.
ARPES measurements in Figs. 1--3 were taken at beamline 5-2 at Stanford Synchrotron Radiation Lightsource (SSRL) with $s$- or $p$-polarized light. Data in Figs. 4--8 were taken at beamline 10.0.1.2 at the Advanced Light Source (ALS) with $p$-polarized light or He1$\alpha$ light (21.2 eV) using a monochromatized helium electron cyclotron resonance (ECR) plasma source. The $(1\overline{1}0)$ plane was the scattering plane for all measurements. At both beamlines, data were taken using a Scienta Omicron DA30L detector. The sample temperature was kept below 20 K and the base pressure during measurement was lower than 3$\times10^{-11}$ Torr. The samples were transferred from the growth systems at UCSB to SSRL and ALS using a custom designed ultrahigh vacuum suitcase with pressure lower than 4$\times10^{-11}$ Torr.

Spin-resolved measurements were performed at beamline 10.0.1.2 using Ferrum spin detectors. The spin texture is measured in three orthogonal directions $(\hat{x},\hat{y},\hat{z})$ which are parallel with our $(k_x, k_y, k_z)$ axes. The spin polarization was calculated from measured spin-resolved energy distribution curves (EDCs) by the usual equation $P=\frac{1}{S}\frac{I_\uparrow-I_\downarrow}{I_\uparrow+I_\downarrow}$, where the Sherman function $S=0.22$. The error bars in polarization are calculated from propagated error in the polarization equation assuming Poisson statistics and neglecting error in the Sherman function. The angular acceptance window and energy window of the spin-resolved measurements is variable, but typically set at 1\textdegree \ and 40 meV. The sign of $\mathrm{P_Y}$ and the Sherman function were validated on bismuth thin film calibration samples.
\begin{figure*}[t]
\includegraphics{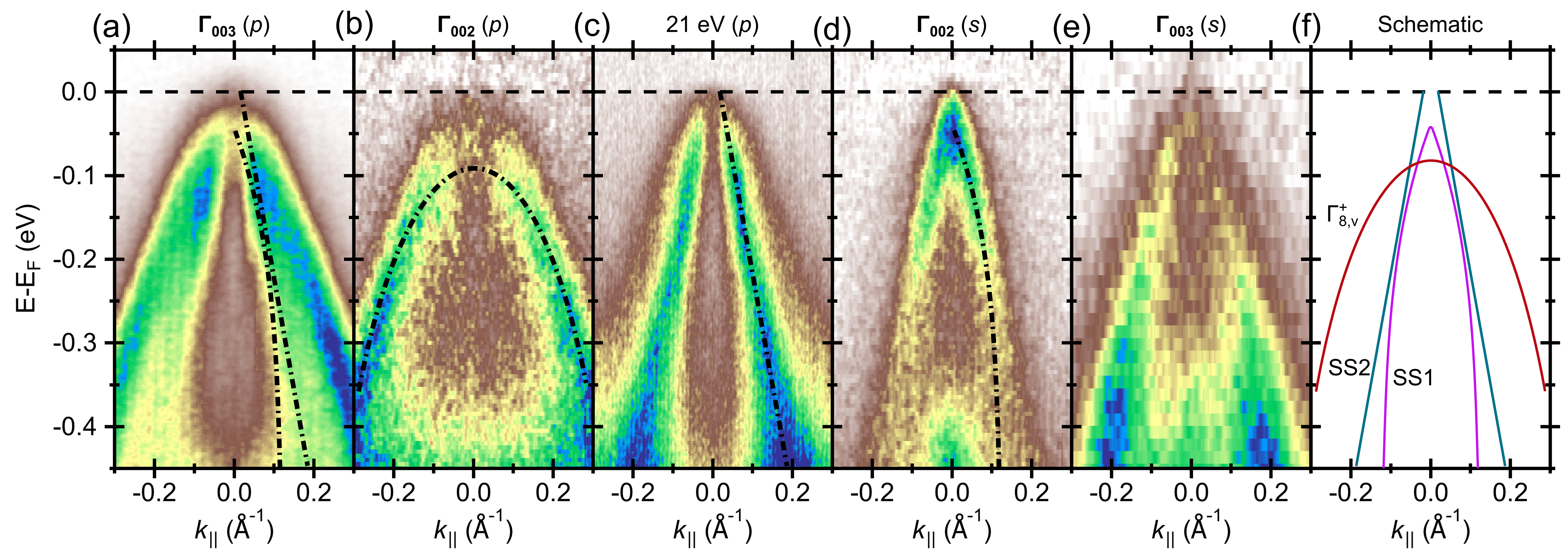}
\caption{\label{Fig2} ARPES measurements close to E$_\textrm{F}$ at (a) \bm{$\Gamma_{003}$} with $p$-polarization (b) \bm{$\Gamma_{002}$} with $p$-polarization (c) $h\nu$=21 eV with $p$-polarization (d) \bm{$\Gamma_{002}$} with $s$-polarization (e) \bm{$\Gamma_{003}$} with $s$-polarization. Guides to the eye are drawn in black. A schematic summarizing the guides to the eye is shown in (f). All measurements are along the $\overline{X}-\overline{\Gamma}-\overline{X}$ direction. \bm{$\Gamma_{003}$} corresponds to $h\nu$=127 eV and  \bm{$\Gamma_{002}$} corresponds to $h\nu$=55.8 eV.}
\end{figure*}
\section{Results and Discussion}
\subsection{Dispersion near bulk $\mathbf{\Gamma}$}\label{ssec:F12}

The bulk-like electronic structure is calculated via tight-binding in Fig. 1(a). The slight compressive strain from growth of $\alpha$-Sn on InSb(001) does not modify the dispersion strongly other than the behavior near the quadratic touching point \cite{Supp}. The band structure is consistent with the expected gapless semiconductor with band inversion. The highest energy band depicted is the $p$-like $\Gamma_{8,c}^+$ band. This band is inverted from its usual character (in the parlance of the Ge band structure, it is the light hole band). It is degenerate at the $\mathbf{\Gamma}$ point with the first valence band, the $\Gamma_{8,v}^+$ band, which has the $p$-like heavy hole (HH) character. The second valence band is the $s$-like $\Gamma_{7}^-$ band, the inverted conduction band. The band inversion in $\alpha$-Sn is between the $\Gamma_{8,c}^+$ and the $\Gamma_{8,v}^+$ band. The third valence band is the $p$-like $\Gamma_{7}^+$ split-off band. The band shapes agree with other calculations \cite{Groves1963,Chen2022,kufner2013}. However, there is still some slight contention in the shape of $\Gamma_{7}^-$ band near its maximum. Some calculations \cite{Barfuss2013,Chen2022, Rohlfing1998,kufner2013} (in addition to Fig. 1(a)) suggest a dimple-like warping away from the parabolic-like dispersion while others \cite{Groves1963, Pollak1970} do not. The presence of this warping has important ramifications for critical point measurements and excitonic effects in this material \cite{Carrasco2018}. The shape of this band is very sensitive to spin-orbit splitting in the system \cite{Carrasco2018}, and thus sensitive to how spin-orbit coupling is taken into account in the given calculation. 

Since the features of interest in $\alpha$-Sn are near the $\mathbf{\Gamma}$ point, we first clarify the band dispersion of both the bulk bands and surface states here, summarized in Figs. 1 (c)--(e). 
These measurements are performed on a 13 BL film which is expected to be in a 3D TI or 2D TI phase \cite{Barfuss2013,Ohtsubo2013,Anh2021,Kufner2014}. In order to deconvolute any matrix element effects in the band structure we investigate the \bm{$\Gamma_{003}$} and \bm{$\Gamma_{002}$} points with both $s$- and $p$-polarized light. Matrix elements are usually defined within both the sudden and the dipole approximations and assuming noninteracting electrons \cite{Sobota2021,Damascelli2004}. 
This results in the one electron dipole matrix element
\begin{equation}
M_{f,i}^\textbf{k}=\langle \phi_f^\textbf{k}|H_{int}|\phi_i^\textbf{k}\rangle
\end{equation}
where $\phi_f^\textbf{k}$ is the final state of the photoelectron, $\phi_i^\textbf{k}$ is the initial state of the photoelectron, and $H_{int}$ is the Hamiltonian representing the electron-photon interaction \cite{Sobota2021}. The intensity of the measurement is proportional to the squared matrix element $I(k,h\nu)\propto|M_{f,i}^\textbf{k}|^2$ while the initial state $\phi_i^\textbf{k}$ is the feature of interest in our measurements \cite{Sobota2021}. The final state $\phi_f^\textbf{k}$ of the photoelectron is usually assumed to be free electron-like with even symmetry; the symmetry of $H_{int}$ varies by the polarization of light being used; the symmetry of the initial state depends on the geometry of the measurement and orbital character of the excited band \cite{Damascelli2004}. From these arguments, tuning the polarization of incident light in the ARPES measurement elucidates the character of a given band \cite{Damascelli2004}. Calculations of the expected effect of matrix element effects in $\alpha$-Sn for $p$- and $s$-polarized illumination are given in Section S1 of the Supplementary Material \cite{Supp}.

The photon energies corresponding to the \bm{$\Gamma_{003}$} and \bm{$\Gamma_{002}$} points are derived from the usual relation 
\begin{equation}\label{eq:innerpot}
k_{z}=\frac{\sqrt{2m_0}}{\hbar}\sqrt{E_k-\frac{\hbar^2}{2m_0}k_{||}^2-V_0}
\end{equation}
using the inner potential model which assumes free electron-like final states. The inner potential, $V_0$, used currently is 5.8 eV \cite{Rogalev2017}, but another derived inner potential of 9.3 eV \cite{Chen2022} does not change the expected dispersion significantly. Since all data in Figs. 1(c)--(e) were taken at the \bm{$\Gamma$} point, the dispersion of the bands in Figs. 1(c)--(e) is identical but matrix element effects shift the relative intensity of certain bands. The linear dichroism between $s$- and $p$-polarized light is not consistent between measurements at \bm{$\Gamma_{003}$} (Fig. 1(c) vs. Fig. 1(d)) and at \bm{$\Gamma_{002}$} (Fig. 1(e) vs. Fig. 1(f)). The main consistency is an emphasis on the $\Gamma_{8,v}^+$ heavy hole band using $p$-polarized light. This indicates that at these photon energies, linear dichroism cannot be treated as a direct probe of orbital characters for the valence bands in $\alpha$-Sn. Since the initial state ($\phi_i^\textbf{k}$) and $H_{int}$ are unchanged, only a change in the final state can then explain the varying matrix element behavior at different photon energies. Final state effects are then likely significant in ARPES measurements of this system.
\begin{figure*}[t]
\includegraphics{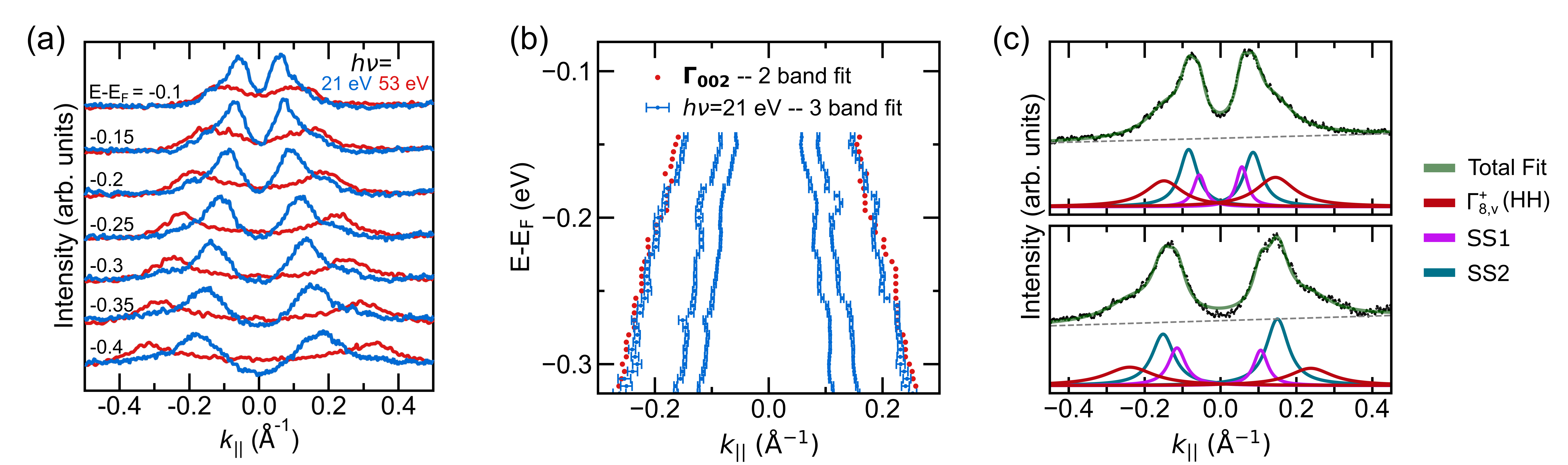}
\caption{\label{Fig3} Detailing the number of surface states in ultrathin $\alpha$-Sn. (a) Representative momentum distribution curves (MDCs) from the $h\nu$=21 eV measurement in red and the \bm{$\Gamma_{002}$} measurement in blue. MDCs are integrated over 10 meV. (b) Fit peak locations for a 3 band fit for the $h\nu$=21 eV measurement (red) and a 2 band fit for the \bm{$\Gamma_{002}$} measurement (blue). (c) Representative three band fits to the MDCs from the $h\nu$=21 eV measurement corresponding to \textit{top}: E--E$_\textrm{F}$=150 meV and \textit{bottom}: E--E$_\textrm{F}$=305 meV.}
\end{figure*}

In Fig. 1(c), most of the valence bands and surface states are visible and agree with their expected dispersion from the calculations. Starting from high binding energy, the split-off band has its maximum 1.1 eV below the Fermi level. There is an additional band inversion in between the split-off band $\Gamma_{7}^+$ and the inverted conduction band $\Gamma_{7}^-$, discussed in detail in Ref. \cite{Rogalev2017} which results in an additional surface state deep below the valence band maximum.
This topological surface state, the \textbf{M}-shaped SS3, (labeled TSS2 in prior work \cite{Rogalev2017,Chen2022}) arises from the split-off band and joins the inverted conduction band near $k_{||}=\pm0.15$ \AA$^{-1}$.  
The maximum of the inverted conduction band is not visible here due to matrix element effects. Past the $k_{||}$ extent of SS3, an additional surface state arises from the inverted conduction band. This surface state then disperses up to the valence band maximum, where it is difficult to distinguish the surface state from the heavy hole band. The heavy hole band ($\Gamma_{8,v}^+$) is then visible as the outermost band. The maximum of the heavy hole band is indicated and notably around 90 meV below the Fermi level. By keeping the measurement at  \bm{$\Gamma_{003}$} and switching to $s$-polarization (Fig. 1(d)), the linear-like surface state (SS2) arising from the inverted conduction band is visible. It disperses up to the Fermi level as indicated. The maximum of the inverted conduction band ($\Gamma_{7}^-$) is also now clearer. No warping of any kind is visible. The difference between the $\Gamma_{7}^-$ and $\Gamma_{8,v}^+$ band maxima is 390 meV in close agreement with the 410 meV measured by other techniques \cite{Carrasco2018}. 

We next move to a measurement with $p$-polarized light at \bm{$\Gamma_{002}$} (Fig. 1(e)). Only the heavy hole band is clear here. A faint intensity corresponding to the inverted conduction band can be seen. Switching to $s$-polarization at the \bm{$\Gamma_{002}$} point (Fig. 1(f)), a surface state again disperses toward the Fermi level from the inverted conduction band ($\Gamma_{7}^-$). This surface state (SS1) is not the same surface state measured in Fig. 1(d) (SS2), investigated in more detail in the next paragraphs. The inverted conduction band is visible, perhaps with a slightly lower effective mass near the band maximum than expected from the \bm{$\Gamma_{003}$} measurement. This could be the result of a narrowing of this band near its maximum as is seen in the analogous HgTe band structure \cite{Vidal2023}.
\begin{figure*}[!ht]
\includegraphics{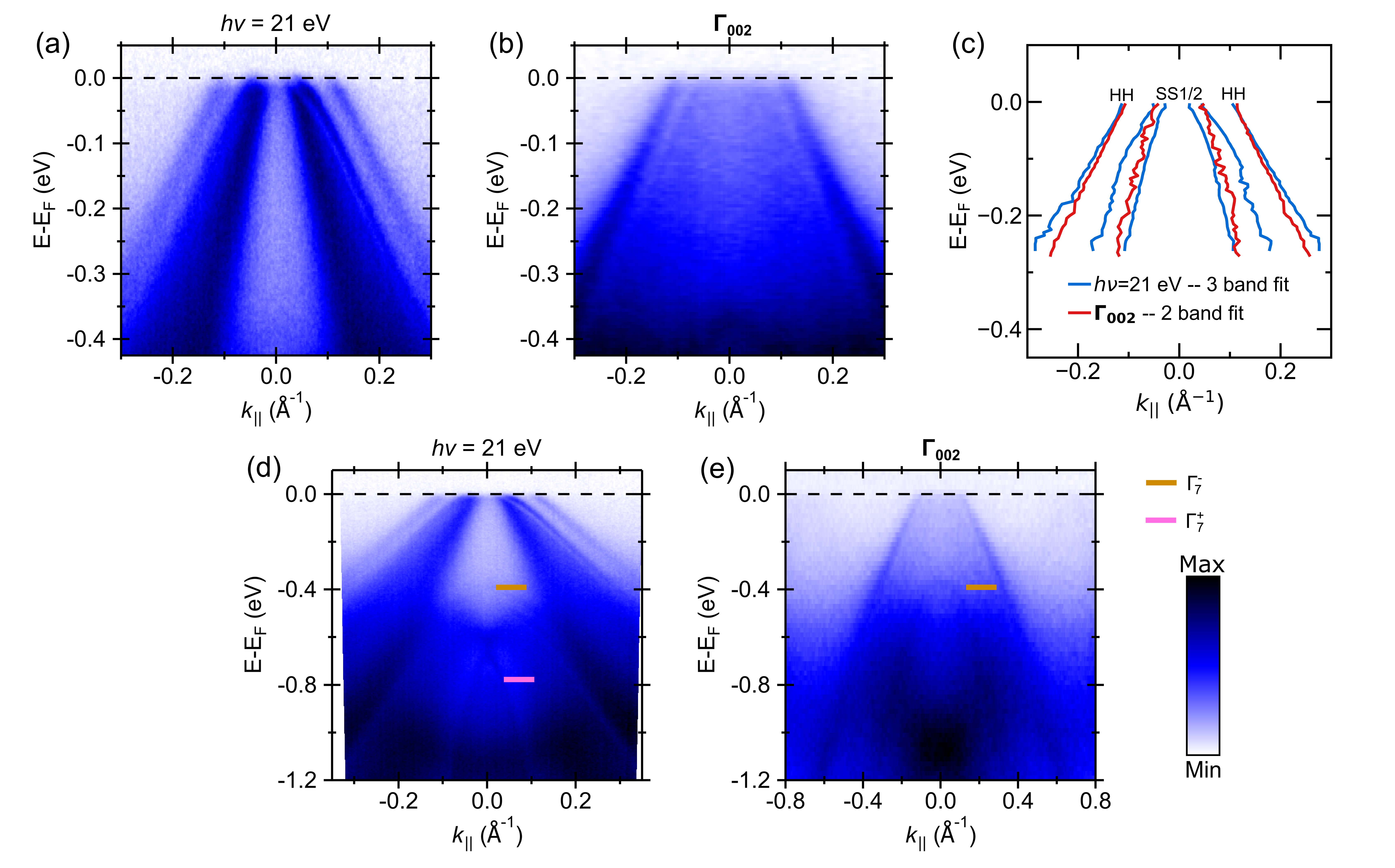}
\caption{\label{Fig4} Detailing the number of surface states in thick $\alpha$-Sn. ARPES measurements at (a) $h\nu$=21 eV and (b) \bm{$\Gamma_{002}$} ($h\nu$=53 eV) for a 400 BL film (400BL-A) along the $\overline{X}-\overline{\Gamma}-\overline{X}$ direction (c) Fit peak locations for a 3 band fit for a $h\nu$=21 eV measurement (red) and a 2 band fit for the \bm{$\Gamma_{002}$} measurement (blue). (d) Wider binding energy range of the measurement in (a). (d) Wider binding energy/momentum range of the measurement in (b). Horizontal colored lines correspond to bulk band maxima. All measurements were performed with $p$-polarized light.}
\end{figure*}

To have a clearer picture of the surface state and bulk dispersion near the Fermi level (and valence band maximum) measurements were performed in the same conditions as Fig. 1 but in a smaller binding energy range (Fig. 2). In Fig. 2(a) the dispersion of the heavy hole band and surface states is shown using $p$-polarized light at \bm{$\Gamma_{003}$}. The guides to the eye shown here are derived from later panels. By keeping the same polarization and switching to \bm{$\Gamma_{002}$} in Fig. 2(b), only the heavy hole band has high intensity. A guide to the eye is shown to indicate the parabolic-like dispersion. The valence band maximum is roughly 90 meV below the Fermi level. A summary of the various band positions derived for the 13 BL film are summarized in Table \ref{tab:table1}.

There is a vanishing intensity of the heavy hole band at \bm{$\Gamma$}, which is visible in films of comparable thickness to ours in Ref. \cite{Rogalev2017} and films of unknown thickness in Ref. \cite{Barfuss2013}. We do not observe the heavy hole band with measurable intensity using $s$-polarized light so it is unclear whether this is a matrix element effect specific to the use of $p$-polarized light, or a feature pertaining directly to the initial state. Changes to the initial state away from the expected parabolic-like band could be from strain-induced distortions calculated via tight-binding and $k\cdot p$ models \cite{Carrasco2018,Rogalev2017}, however in our calculations we find--for the moderate strains induced by the -0.15\% strain in this system--the bowing is quite small compared to experiment (Section S1 of Supplemental Material) \cite{Supp}. In addition in our model we partially replicate this missing intensity feature using $p$-polarized light  \cite{Supp}. Furthermore, quantum well states derived from the heavy hole band (discussed further in Section \ref{ssec:confinement}) retain this missing intensity at \bm{$\overline{\Gamma}$}. 

In order to investigate the surface state structure of $\alpha$-Sn in more detail, photon energies in the range of 17-23 eV are frequently used. We use $h\nu$=21 eV (Fig. 2(c)) in order to better compare to reports in the literature at similar photon energies. This photon energy results in a $k_z$ value either halfway between the \bm{$\Gamma_{001}$} and \bm{$Z$} points (using an inner potential of 5.8 eV \cite{Rogalev2017}) or an additional $\sim$10\% closer to the \bm{$Z$} point (using an inner potential of 9.3 eV \cite{Chen2022}). Referencing the tight-binding calculations in Fig. 1(a) and the heavy hole maximum (Table 1), the highest lying bulk band in a $h\nu$=21 eV measurement should be the heavy hole band ($\Gamma_{8,v}^+$), which should have its maximum $\sim$1.5 eV below the Fermi level. As Fig. 2(c) only probes 500 meV below the Fermi level, no bulk bands should be visible in this measurement. 
This measurement is dominated by a linear surface state (SS2) which has a crossing $\sim$37 meV above the Fermi level.
In Fig. 2(d) at \bm{$\Gamma_{002}$} with $s$-polarization, a different surface state (SS1) has significant intensity. SS1 has a crossing $\sim$50 meV below the Fermi level (and thus $\sim$40 meV above the valence band maximum). We do not see any evidence of an upper branch to this surface state as would be expected in a Dirac-like topological surface state, however this could arise from the proximity of the crossing to the Fermi level. There does not appear to be any significant anticrossing-like behavior as the surface states disperse through the heavy hole band. Finally, at the \bm{$\Gamma_{003}$} point with $s$-polarization (Fig. 2(e)), the linear SS2 can be seen with a dispersion closely matching the guide to the eye in Fig. 2(c).
\begin{table}
\caption{\label{tab:table1}
Energies corresponding to band maxima at \bm{$\Gamma$} (for bulk bands) and surface state crossings (for surface states) referenced to E$_\textrm{F}$ in the 13 BL film.}
\renewcommand{\arraystretch}{1.5}%
\begin{ruledtabular}
\begin{tabular}{c|cccccc}
 &
\textrm{SS1}&
\textrm{SS2 (fit)}&
\textrm{SS3}&
$\Gamma_{8,v}^+$&
$\Gamma_{7}^-$&
$\Gamma_{7}^+$\\
\colrule
E--E$_\textrm{F}$ (eV)&-0.05&0.04&-0.72&-0.09&-0.48&-1.17\\
\end{tabular}
\end{ruledtabular}
\end{table}
A schematic summarizing the dispersion of the bands observed in 13 BL $\alpha$-Sn is given in Fig. 2(f). Importantly, we observe only two surface states in ultrathin $\alpha$-Sn(001) which disperse into the bulk band gap transitioning from surface resonances to true surface states. Of these two surface states, the dispersion of SS2 is more consistent with that of a Dirac-like topological surface state.

\subsection{Clarification to the number of surface states in $\alpha$-Sn(001) films}\label{ssec:fitting}
Generally VUV measurements far from the \bm{$\Gamma$} point in $\alpha$-Sn(001) show evidence of three states which are associated with either three surface states \cite{Chen2022} or two surface states and a bulk band \cite{Scholz2018}. These observations span a large range of film thicknesses such that quantum confinement effects are likely not in play. While three bands are visible in our \bm{$\Gamma_{003}$} measurement with $p$-polarization in Fig. 2(a) (SS1, SS2, and HH), only SS2 is readily visible with $p$-polarization in Fig. 2(c). A series of momentum distribution curves (MDCs) taken from the measurement in Fig. 2(c) are plotted in Fig. 3(a). The lineshapes are consistent with those presented in Ref. \cite{Scholz2018} and indicative of three bands. The two innermost states correspond to SS1 and SS2, but the identification of the outermost state is unclear. These MDCs are then plotted against MDCs taken from Fig. 2(b) where primary intensity is due to the heavy hole band. The outermost peak in the MDCs from the \bm{$\Gamma_{002}$} measurement, which corresponds to the heavy hole band, lines up with the outermost peak in the MDCs from the $h\nu$=21 eV measurement. This indicates the third and outermost peak seen in these VUV measurements could actually be the heavy hole band.

This is verified by performing a three band fit to the MDCs of the $h\nu$=21 eV data and a two band fit to the MDCs of the \bm{$\Gamma_{002}$} data, summarized in Fig. 3(b). The peaks are fit to a Voigt lineshape where the Gaussian component is fixed at the experimental momentum resolution. The fit is constrained such that there is approximate symmetry of the peak locations, amplitudes, and FWHMs across $\overline{\Gamma}$. The error bars for the peak locations in the three band fit are typical for the fits to peak location in the rest of this work. We exclude the error bars from other plots for clarity, as they do not have a meaningful effect on our conclusions. The outermost state in the two fits in Fig. 3(b) lines up almost exactly, confirming that the three states seen in our VUV measurement are SS1, SS2, and the HH band. Representative three band fits to the $h\nu$=21 eV measurement, showing its suitability for the given system are given in Fig. 3(c) for 150 meV and 305 meV binding energy. While these identifications have been confirmed in ultrathin films, we would like to show this is a global effect for the $\alpha$-Sn(001) surface independent of film thickness.

In order to clarify this point further, 400 BL $\alpha$-Sn films are investigated. At this thickness, there should not be significant confinement effects and the films are expected to be in the 3D DSM phase \cite{deCoster2018,Anh2021,Chen2022}. The measurement with $p$-polarization at $h\nu$=21 eV is shown in Fig. 4(a)  on 400 BL-A where the presence of three tightly dispersing states can now be clearly seen, all crossing the Fermi level. The surface state crossings and valence band maximum are 50--100 meV above E$_\mathrm{F}$. In Fig. 4(b) the dispersion at the \bm{$\Gamma_{002}$} point is shown. The heavy hole band is clearly resolved, while the surface states are broad and difficult to observe. Fits are performed on the MDCs of these measurements to more precisely compare their shapes. A three band fit is performed for the $h\nu$=21 eV measurement. We could not perform a robust three band fit to the \bm{$\Gamma_{002}$} data due to the lack of sharpness of the bands. Instead, we perform a two band fit with the expectation one of the bands will essentially average SS1 and SS2. The fit dispersions are summarized in Fig. 4(c) where this assumption is shown to be correct. The outermost band in both fits lie almost directly on top of each other. The measureed constant energy contours of the heavy hole band also show good agreement with tight-binding calculations of the predicted constant energy contours using $p$-polarized light \cite{Supp}. Thus, the outermost state measured in ARPES of both thin and thick films of $\alpha$-Sn(001) using VUV light is likely the heavy hole band at \bm{$\Gamma$} rather than a third surface state, even though the $k_z$ value estimated using the inner potential model disagrees quite strongly with this explanation.

If the $h\nu$=21 eV measurement is sampling the heavy hole at $k_z=0\times\frac{4\pi}{c}$, ostensibly it should sample the other bulk bands at $k_z=0\times\frac{4\pi}{c}$ as well. A wider energy range of the $h\nu$=21 eV measurement is shown in Fig. 4(d). The maximum of the inverted conduction band is hazily visible 390 meV below E$_\mathrm{F}$. This location agrees with a more direct measurement of the $\Gamma_{7}^-$ band maximum at the \bm{$\Gamma_{002}$} point in Fig. 4(e). Using the $\Gamma_{7}^-$--$\Gamma_{8,v}^+$ spacing derived earlier for the 13 BL film, the valence band maximum would be only 20 meV above the Fermi level. Likewise, the inverted conduction band ($\Gamma_{7}^-$) is visible with its expected \bm{$\Gamma$} point dispersion. The node of SS3 is approximately 630 meV below the Fermi level. A gap in intensity can be seen in between the node of SS3 and the maximum of the split-off band ($\Gamma_{7}^+$). The maximum of the split-off band is 385 meV below the maximum of the $\Gamma_{7}^-$ band. The dispersion of the $\Gamma_{7}^-$ band agrees with that shown in Fig. 1(d), appearing parabolic-like with minimal warping.

So far we have only investigated the appearance of the unexpected \bm{$\Gamma$} point dispersion using $h\nu$=21 eV along the $\overline{X}-\overline{\Gamma}-\overline{X}$ direction. In Fig. 5, constant energy contours of 400 BL-B $\alpha$-Sn are plotted. In Fig. 5(a) for $h\nu$=21 eV, SS1 and SS2 have the highest intensity. SS1 appears to nearly isotropic. The exact dispersion of SS2 is difficult to resolve, but it appears less isotropic with maximum intensity along the $\overline{M}-\overline{\Gamma}-\overline{M}$ direction (four-fold symmetry). Finally the heavy hole band appears as a broad background at higher $k$ than SS1 and SS2. It also appears four-fold symmetric, as would be expected for the heavy hole band. In Fig. 5(b) for the \bm{$\Gamma_{002}$} measurement, only the heavy hole band is visible, matching the broad background in Fig. 5(a). This contour retains the expected four-fold symmetry. Finally in the measurement at \bm{$\Gamma_{003}$} (Fig. 4(c)), SS1 and SS2 have the same contours as in Fig. 4(a), but here the intensity of the heavy hole band is greater. The contour of the heavy hole band agrees with the \bm{$\Gamma_{002}$} and $h\nu$=21 eV measurements as well. The measured contours for the heavy hole band in Fig. 5 also agree nicely with the calculated contours for the heavy hole band in Section S1 of the Supplementary Material \cite{Supp}. Thus, the ARPES measurements of $\alpha$-Sn(001) with VUV light deviate strongly from the inner potential model using the generally accepted values of the inner potential.

What is not yet clear is where the discrepancy with the inner potential model arises. The inner potential $V_0$ (Eq.~(\ref{eq:innerpot})), is defined as $V_0=E_0-e\Phi$, where $E_0$ is the zero energy of the (assumed) free electron final band with respect to the valence band maximum of the semiconductor and $\Phi$ is the work function of the semiconductor \cite{Leckey1992}. The simplest explanation for the discrepancy we measure is that the constant inner potentials found in prior work of 5.8 eV \cite{Rogalev2017} (measured using soft X-ray light) or 9.3 eV \cite{Chen2022} (measured using extreme UV light) are incorrect. However, our results in Figs. 1 \& 2 agree nicely with these computed inner potentials (and our own photon energy dependence measurements). Furthermore, to produce a \bm{$\Gamma$} at $h\nu$=21 eV, $V_0$ would need to be near 40 eV, an unphysical value. 

One potential explanation is that the assumption of a free electron final state with constant inner potential fails at these low photon energies. In photon energy dependent measurements of $\alpha$-Sn in the VUV range in the past, a $h\nu$ dependent inner potential term was adopted such that the final state is parabolic with an effective mass of 0.22$m_0$, although this approach was \textit{ad hoc} \cite{Middelmann1987}. 
\begin{figure*}[!ht]
\includegraphics{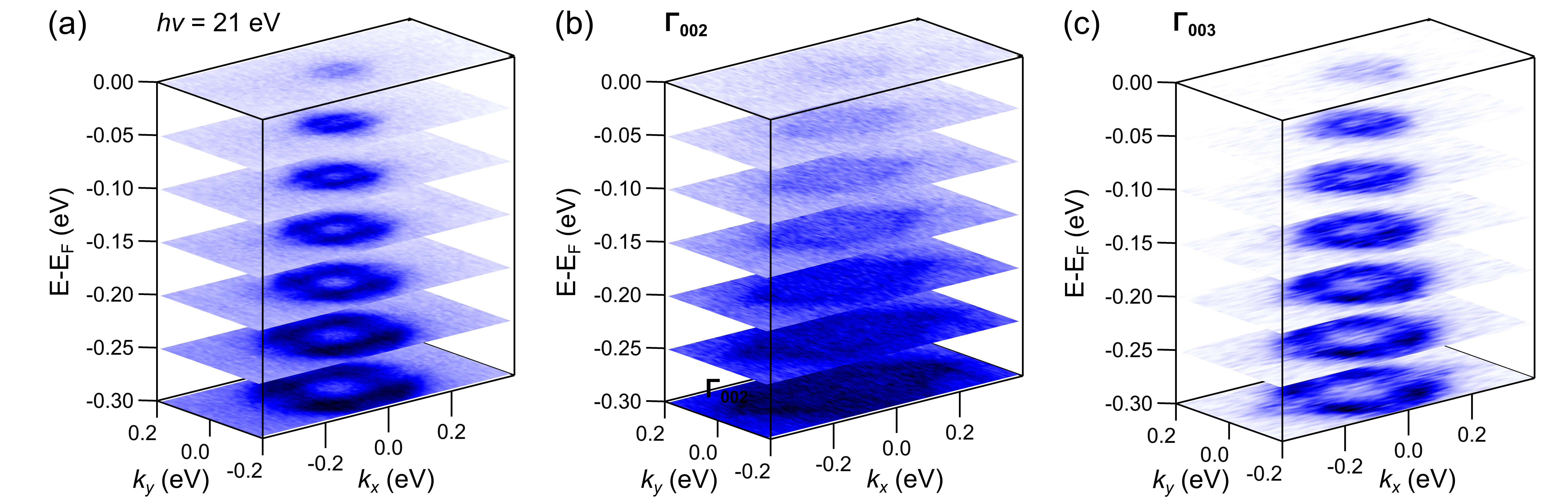}
\caption{\label{Fig5} Constant energy countours taken at (a) $h\nu$=21 eV, (b) \bm{$\Gamma_{002}$} ($h\nu$=53 eV) and (c) \bm{$\Gamma_{003}$} ($h\nu$=127 eV)for a 400 BL film (400BL-B) with $p$-polarized light. Contours were integrated over a 10 meV window.}
\end{figure*}
It has been well-established that for lower photon energy ARPES measurements of GaAs(001) ($h\nu<$ 50 eV) the expected dispersion assuming a free electron final state with constant inner potential varies in its suitability as a function of photon energy and as a function of the band in the initial state, while for GaAs(110) the constant inner potential model can be quite robust \cite{Leckey1992}. 
Deviations from this model vary by material, crystalline orientation, band of interest, and photon energy range used. 
Failures are then not unexpected and disagreements between measured band structure via ARPES and calculated band structure could just as likely be from deviations in the final states as deviations in the initial states \cite{Leckey1992}. 
In fact, deviations from the free electron final state were already observed in $\alpha$-Sn via the dichroism measurements in Section \ref{ssec:F12}. 

Even assuming the inner potential model holds as expected, prominent $k_z$ broadening can be present in low photon energy measurements. $k_z$ broadening is caused by the Heisenberg uncertainty principle between $k_z$ and the inelastic mean free path  \cite{Sobota2021,Zhang2022ARp}. 
The typical constant photon energy $E-k_{||}$ measurement is then not truly a slice at some given $k_z$ value (determined by Eq.~(\ref{eq:innerpot})), but actually integrated over some $\Delta k_z$ inversely related to the inelastic mean free path (IMFP) of the photoelectrons \cite{Strocov2003}. 
If the broadening (i.e. total $k_z$ sampled) is very large, band features could show up in unexpected locations. The IMFP for photoelectrons near the Fermi surface in the $h\nu$=21 eV measurement (calculated via a modified fit to the universal curve \cite{Seah1979}) is 4.3 \AA, close to the 5.0 \AA\ IMFP (calculated by the same method) for the bulk-sensitive $h\nu$=352 eV measurements in prior work \cite{Rogalev2017}. 

Another possible explanation for the presence of these states is that they are indirect transitions. When the energy of the final state is low, the effect of the crystal potential cannot be neglected \cite{Chiang2001}. This perturbation can result in a dispersion expected from a high symmetry point to be seen at photon energies that do not correspond to that symmetry point \cite{Chiang2001}. In the similar system of CdTe(111) only indirect transitions are visible in ARPES measurements using photon energies in the 19-30 eV range; these indirect transitions reflect a Fourier sum of final states arising from high symmetry initial states corresponding to $k_z$=0 and 0.5 (center and edge of the Brillouin zone)\cite{Ren2015,Jardine2023}. These indirect transitions show no photon energy dependence, the opposite of that which would normally be expected from direct bulk band transitions. If the presence of the  \bm{$\Gamma$} point like bulk band dispersion in the 21 eV  measurements presented here is rooted in indirect transitions, any ARPES measurements in the VUV range where these indirect transitions are present would then be expected to show the HH band in close proximity to the set of two surface states.

Neither the situation of indirect transitions nor a failure of the free electron final state assumption can be isolated in our current measurements, but both are reasonable explanations. Our results emphasize that the final states of $\alpha$-Sn should be investigated in more detail such that better convergence can be reached between the predicted and measured initial state electronic structure of $\alpha$-Sn. 

\subsection{Spin-polarized ARPES of the surface states near the valence band maximum in $\alpha$-Sn(001)}\label{ssec:spin}
\begin{figure*}[!ht]
\includegraphics{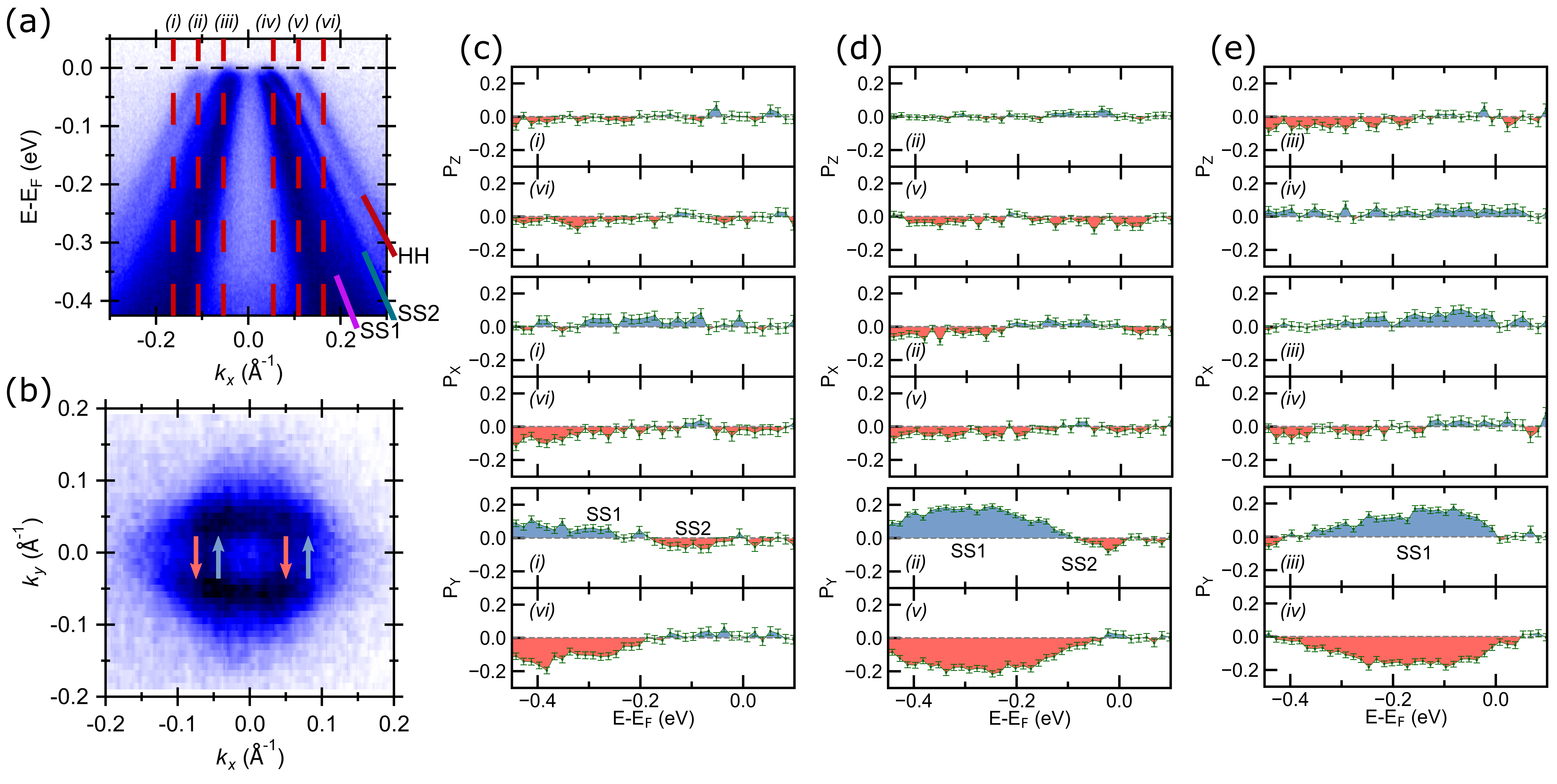}
\caption{\label{Fig6} Determination of the spin texture of the surface states. (a) ARPES measurement of 400 BL-A with the energy distribution curves (EDCs) along which spin polarization was measured indicated with the red dashed line. (b) A constant energy countour taken from Fig. 5(a) at E--E$_\textrm{F}$. Spin polarization measured in ($\hat{x}$,$\hat{y}$,$\hat{z}$) at (c) $k_{x}=\pm$0.163 \AA$^{-1}$, (b) $k_{x}=\pm$0.109 \AA$^{-1}$, and (c) $k_{x}=\pm$0.054 \AA$^{-1}$. All measurements were made at $h\nu$=21 eV with $p$-polarization. Raw SARPES data is given in Section S3 of the Supplemental Material \cite{Supp}.}
\end{figure*}
To help identify the character of the surface states (SS1 \& SS2) we performed spin-resolved ARPES measurements on the same sample 400 BL-A. Spin polarization was investigated in the $(\hat{x},\hat{y},\hat{z})$ directions summarized in Fig. 6. The EDCs along which spin polarization was measured are indicated in Fig. 6(a). A constant energy contour corresponding to E--E$_\textrm{F}=-0.1$ eV in Fig. 5(b) is shown in Fig. 6(b). At large $k_x$ (Fig. 6(c)) there is negligible spin polarization in the $\hat{z}$ direction. There is a very slight spin polarization at high binding energy at positive $k_x$, but the spin polarization does not obey time-reversal symmetry so it is unlikely to be an initial state effect. At smaller $k_x$ (Fig. 6(d)), there is no spin polarization in either the $\hat{x}$ or $\hat{z}$ directions. At the smallest $k_x$ (Fig. 6(e)), there is no polarization in the $\hat{z}$ direction. There is a slight increase of spin polarization near E$_\textrm{F}$ at negative $k_x$, but as before it does not obey time-reversal symmetry. Since P$_\textrm{X}$ and P$_\textrm{Z}$ are zero at all points measured, any non-zero P$_\textrm{Y}$ then results in the ideal helical spin-momentum locked spin texture of the surface state as expected if it were a Dirac-like topological surface state or Rashba-split surface state.

Indeed in P$_\textrm{Y}$, a finite spin polarization is measured. Rather than the one spin state measured in past work \cite{Barfuss2013,Ohtsubo2013,Scholz2018}, we see evidence of two spin states. At $k_{x}=\pm$0.163 \AA$^{-1}$ there is a state with P$_\textrm{Y}$$\approx$10\% polarization peaked 400 meV below E$_\textrm{F}$, while the second spin state -- with opposite spin polarization -- is peaked 100 meV below the Fermi level with a small polarization. Both spin states obey time reversal symmetry. At $k_{x}=\pm$0.109 \AA$^{-1}$ the spin polarization of the lower energy state is much larger, around 20\%, and peaks near 300 meV. The second spin state shifts in binding energy by a similar amount and retains its much smaller spin polarization. Finally at $k_{x}=\pm$0.054 \AA$^{-1}$, where the heavy hole band is above E$_\textrm{F}$, the spin polarization due to the lower binding energy state is no longer visible. There appears to be a constant shift in P$_\textrm{Y}$ by about $-5\%$ across these measurements ($k$-independent), which is likely due to the spin-dependent photoemission matrix elements (SMEs) that have been observed in Bi$_2$Se$_3$ \cite{Jozwiak2011,Jozwiak2013}. In addition, the measured P$_\textrm{Y}$ decreases at large $k_x$ values (See Section S3 of \cite{Supp}). For now, we attribute the larger spin polarization peak to SS1 and the smaller spin polarization peak to SS2. This attribution (that the smaller peak corresponds to SS2 rather than the heavy hole band) is shown more rigorously in the next paragraphs. SS1 and SS2 are thus spin-polarized, having the ideal orthogonal spin-momentum locking and opposite helicities (Fig. 6(b)). 

The magnitude of the spin polarization in SS1 is roughly half that of the spin polarization measured for $\alpha$-Sn(001) in Barfuss \textit{et al.} \cite{Barfuss2013} and roughly equal to that measured in Scholz \textit{et al.} \cite{Scholz2018}, but the value of the polarization measured in  Ohtsubo \textit{et al.} \cite{Ohtsubo2013} is unknown. The three prior SARPES measurements give overall similar results where only one spin state was visible and said spin state was associated with a Dirac-like topological surface state. The sign of the helicity of SS1 measured in Fig. 6 agrees with that found in Ohtsubo \textit{et al.} \cite{Ohtsubo2013}, but has opposite helicity to that found in Barfuss \textit{et al.} \cite{Barfuss2013} and Scholz \textit{et al.} \cite{Scholz2018}. While our work, Barfuss \textit{et al.} \cite{Barfuss2013}, and Ohtsubo \textit{et al.} \cite{Ohtsubo2013} all report a perfect spin-momentum locking in SS1, Scholz \textit{et al.} \cite{Scholz2018} reports some canting with a finite spin polarization in the $\hat{x}$ direction. One possible root of these discrepancies is the experimental geometry. The polarization of light has been found to be a tool to manipulate the spin vector of the topological surface states in Bi$_2$Se$_3$-based systems \cite{Jozwiak2013,Kuroda2022}. In Ohtsubo \textit{et al.} \cite{Ohtsubo2013}, which measures an equivalent helicity of SS1 to that in Fig. 6, the experimental geometry is equivalent to ours other than a $C_2(z)$ rotation of the two-fold symmetric substrate. This should not have a large effect on these measurements as the measured electronic structure of the $\alpha$-Sn film is four-fold symmetric (Fig. 5). Moreover, modification of the SS1 spin texture by $p$-polarized light (as used here), should it follow the mechanism found in Ref. \cite{Jozwiak2013} and Ref. \cite{DamascelliBSspin} for Bi$_2$Se$_3$-based systems, should show strong deviations from orthogonal spin-momentum locking at $k$ other than those measured here. However, further measurements away from high symmetry lines found the ideal orthogonal spin-momentum locking (see Section S4 of \cite{Supp}) indicating that the polarization of light does not have a strong effect on the measured spin texture of these samples.
The spin canting in the $\hat{x}$ direction in Scholz \textit{et al.} \cite{Scholz2018} could also potentially result from interference of spin-momentum locked textures of SS1 and SS2 if the intrinsic linewidths of these states overlap \cite{Meier2011}. 

Finally, the prior SARPES measurements of $\alpha$-Sn(001), did not find the presence of any spin polarization in SS2. This discrepancy between our measurements and the literature could arise from the high energy resolution of our measurements which is generally 3$\times$ that of other work. In addition, in all prior spin-resolved ARPES measurements the surface and/or the bulk of $\alpha$-Sn is doped with Te or Bi to improve the surface quality and electron-dope the (generally degenerately $p$-type doped films) films. These treatments could strongly modify the measured surface electronic structure and have already been seen to renormalize the velocity of the surface states \cite{engel2023growth,Madarevic2020,Chen2022}. Impurities on the surface have also been calculated to potentially lead to a non-orthogonal spin-momentum locking\cite{CantingSpin}. The films used in our study are free from these treatments, and thus the effects measured here are intrinsic to $\alpha$-Sn(001).

Now we investigate our attribution of the spin polarization as arising from the initial states of SS1 and SS2 by varying both the polarization of light and the resolution of the measurement. These measurements were performed at $k_{x}=\pm$0.109 \AA$^{-1}$ where the intensity and bandwidth of the SS2 peak in P$_\textrm{Y}$ is maximal. A representative $h\nu$=21 eV ARPES measurement for these samples is shown in Fig. 7(a). The resolutions for the following spin-resolved measurements are indicated in Fig. 7(a) as well and correspond to an energy resolution better than 100 meV for Figs. 7(b,c), 63 meV for Figs. 7(d,e), and 33 meV for Figs. 7(f,g) and an angular acceptance of $\pm$0.5\textdegree, $\pm$0.5\textdegree and $\pm$0.25\textdegree, respectively. 
\begin{figure}[!ht]
\includegraphics{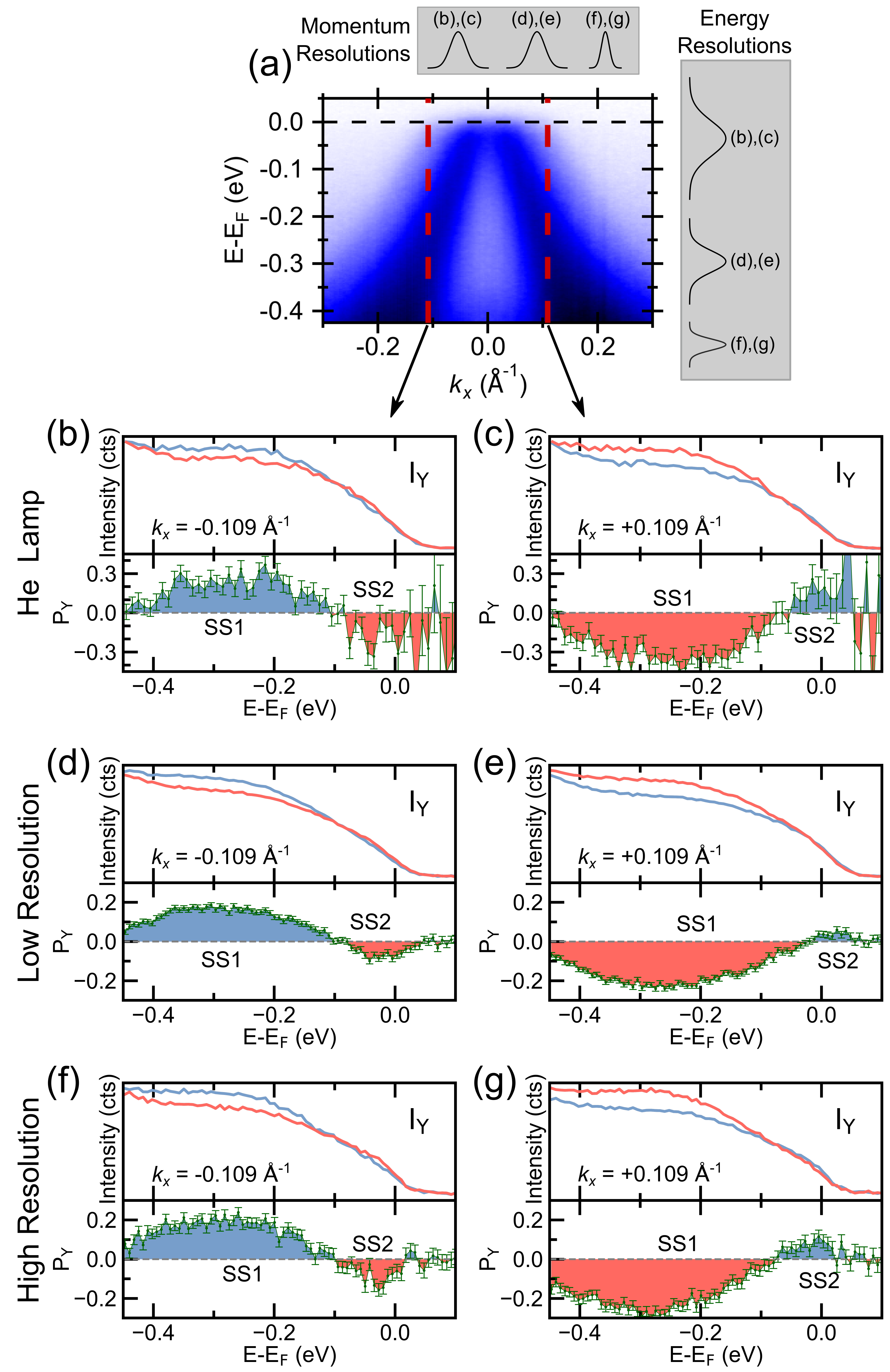}
\caption{\label{Fig7} Verifying spin polarization in SS1 and SS2. (a) ARPES measurement of 400 BL-D using $p$-polarized light at $h\nu$=21 eV with the EDCs along which spin polarization was measured indicated with the red dashed line. P$_\textrm{Y}$ measurement performed using He1$\alpha$ light on 400 BL-C at (b) $k_{x}=-0.109$ \AA$^{-1}$ and (c) $k_{x}=+0.109$ \AA$^{-1}$A. P$_\textrm{Y}$ measurements on 400BL-D using $p$-polarized synchrotron light at $h\nu$=21 eV taken at (d) $k_{x}=-0.109$ \AA$^{-1}$ and (e) $k_{x}=0.109$ with low energy/momentum resolution. P$_\textrm{Y}$ measurements taken immediately afterward on 400 BL-D at (f) $k_{x}=-0.109$ \AA$^{-1}$ and (g) $k_{x}=+0.109$ with high energy/momentum resolution.}
\end{figure}

Spin-resolved photoemission measurements can sometimes measure a finite spin polarization for a feature that does not have a spin-polarized initial state \cite{Osterwalder2006}. These features generally arise from strong spin-orbit coupling related matrix elements and the details of the final state, but other mechanisms such as spin-dependent emission and spin-dependent photoelectron transport can have an effect as well \cite{Osterwalder2006,Heinzmann2012}. Figs. 7(b,c) probe the spin polarization of SS1 and SS2 using the He1$\alpha$ line isolated from the emitted spectrum of a helium ECR plasma source with a monochromator. This light is not entirely unpolarized as a) the light generated in an ECR plasma could have some slight polarization \cite{Iwamae2005} and b) passing the light through a monochromator will partially polarize it. The geometry of the monochromator results in partially linearly polarized light roughly 45\textdegree from the incidence plane. Even so, the total $p$-polarization using the He1$\alpha$ line should be less than that using the $p$-polarized synchrotron light and any features induced by the use of $p$-polarized should show a strong reduction. The matrix element effect of partially polarized light is a combination of the matrix element effects for the unpolarized and polarized contributions of the light \cite{Cherepkov1979}. Neither of these matrix element effects should obey time reversal symmetry \cite{Cherepkov1979}. In Figs. 7(b,c) the spin polarization of SS1 and SS2 are still clear and show time reversal symmetry; The use of the He1$\alpha$ line actually increases the maximum spin polarization of each state by $\sim$50\%.

Since the heavy hole band (the outermost band measured at $h\nu$=21 eV, discussed in Section \ref{ssec:fitting}) is spin degenerate it cannot produce this measured spin polarization; we find that the low binding energy spin polarization is likely from SS2. The helicities of the spin textures of SS1 and SS2 agree with that measured using $p$-polarized light —- the spin texture depicted in Fig. 6(b) (opposite to that typically measured for Dirac-like surface states) seems to then truly be the initial state spin texture, unmodified by matrix element effects from the use of $p$-polarized light, in agreement with the discussion in Section S4 of the Supplemental Material \cite{Supp}.

In these measurements and those in Fig. 6, there is no region of null spin polarization in between the SS1 and SS2 peaks. There is therefore an overlap in the measured spin polarization from the Gaussian resolution function of the measurement. The dependence on the measured polarization on experimental resolution is investigated further using synchrotron light. Referencing Fig. 5(a), at $k_{x}=\pm$0.109 \AA$^{-1}$ (neglecting the angular acceptance window of the spin-resolved measurement) there is an energy spacing of 150 meV between SS1 and SS2 and 230 meV between SS1 and the HH band. Since the distance between SS1 and the HH band is so much larger than the energy resolutions used in Figs. 7(d--g), this change in energy resolution should not have an effect on the measured spin polarization. The angular acceptance window and thus the momentum resolution was varied in this measurement as well. Treating SS1, SS2, and the HH as parallel, the spacing between SS1 and SS2 is roughly 0.05 \AA$^{-1}$, while the spacing between SS1 and the HH band is roughly 0.1 \AA$^{-1}$. The angular acceptance windows correspond to a momentum resolution of 0.036 \AA$^{-1}$ and 0.018 \AA$^{-1}$\ for Figs. 7(d,e) and Figs. 7(f,g), respectively. Similar to the argument with energy resolution, the SS1-HH momentum spacing is large enough that the change in momentum resolution between the two measurements should have a negligible effect if the HH band gave rise to the spin polarization nearer to the Fermi level.

In Figs. 7(d,e) spin polarization data at moderate energy and momentum resolution is shown. At negative $k_x$ the inflection point between the spin states is 80 meV below E$_\textrm{F}$ and at positive $k_x$ it is 20 meV below E$_\textrm{F}$. By improving the resolution of the measurement, the reduced breadth of the resolution function should reduce overlap in the measured spin polarization. High energy/angle resolution spin-resolved measurements at the same $k_x$ values on the same sample, immediately after the prior measurements, are shown in Figs. 7(f,g). Here at negative $k_x$ the spin inflection point is 100 meV below E$_\textrm{F}$, while at positive $k_x$ it is 75 meV below E$_\textrm{F}$. At negative $k_x$ the inflection point shifts 20 meV downward with improved resolution, while at positive $k_x$ the inflection point shifts 55 meV downward. The discrepancy between these values is likely from the constant offset of -5\% in P$_\textrm{Y}$ or a slight asymmetry in the experimental geometry. In addition, the spin polarization of SS1 and SS2 is increased. These large changes do not reflect the small changes expected if the spin polarization near E$_\textrm{F}$ is from the heavy hole band: the measured spin polarization arises from SS1 and SS2 rather than SS1 and the heavy hole band ($\Gamma_{8,v}^+$).
\begin{figure}[!ht]
\includegraphics{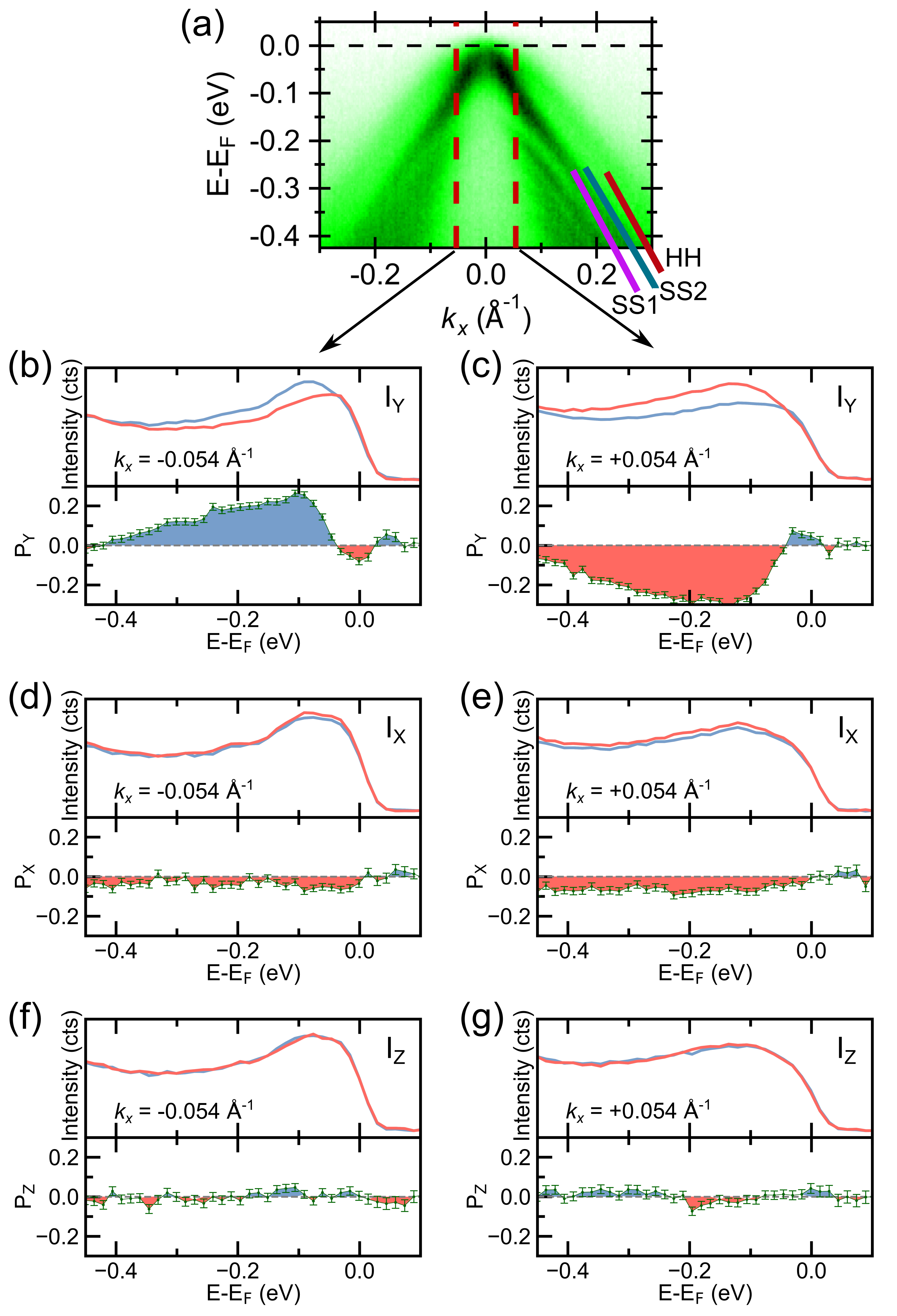}
\caption{\label{Fig8} Spin polarization measurements on 50 BL $\alpha$-Sn(001). (a) ARPES measurement with the EDCs along which spin polarization was measured indicated with the red dashed line. P$_\textrm{Y}$ measurement at (b) $k_{x}=-0.054$ \AA$^{-1}$ and (c) $k_{x}=+0.054$ \AA$^{-1}$A. P$_\textrm{X}$ measurement at (d) $k_{x}=-0.054$ \AA$^{-1}$ and (e) $k_{x}=+0.054$ \AA$^{-1}$A. P$_\textrm{Z}$ measurement at (f) $k_{x}=-0.054$ \AA$^{-1}$ and (g) $k_{x}=+0.054$ \AA$^{-1}$. All measurements were made at $h\nu$=21 eV with $p$-polarization.}
\end{figure}
\subsection{The effect of confinement on spin polarization}\label{ssec:confinement}
Most studies of $\alpha$-Sn with topology in mind have been on films much thinner than 400 BL where quantum confinement plays a stronger role. We investigate this effect in a $h\nu$=21 eV ARPES measurement of 50 BL of $\alpha$-Sn (Fig. 8(a)). SS1 is clearly visible here, but SS2 and the HH band are not. The film is slightly electron-doped with respect to the 400 BL films. The linewidth of SS1 in this film is comparable to the linewidth of SS1 in the 13 BL film (Fig. 2), but noticeably narrower than the linewidth in the 400 BL film (Fig. 4). The linewidth of SS1 in the thin films is also relatively constant with binding energy (consistent with prior work \cite{Scholz2018}), while the linewidth of SS1 in the 400 BL film strongly increases with larger binding energy. This linewidth dependence could be due to enhanced electron-electron scattering, electron-hole scattering, impurity scattering, or a rougher surface \cite{Baumberger2001,Park2010,Ingle2005} in the 400 BL film. 400 BL films are expected to be a 3D DSM compared to a 3D TI or 2D TI for the 13 and 50 BL films \cite{Anh2021,deCoster2018,Kufner2014,Chen2022}.

A series of quantum well states is apparent with the highest lying subband 200 meV below the Fermi level. These quantum well states have the same missing intensity at the \bm{$\Gamma$} point observed in 13 BL films (Figs. 1 and 2). This indicates the missing intensity is intrinsic to the heavy hole band. However, this feature in the quantum well states still does not clarify whether the missing intensity is due to a matrix element effect or a hybridization effect -- the quantum well states should have the same orbital character as the band it is derived from. However, tight-binding calculations including matrix elements of the photoemission process imply that the missing itensity is a matrix element effect \cite{Supp}. The inheritance of this missing intensity from the heavy hole band also confirms that these are in fact quantum well states, as opposed to Volkov-Pankratov states, which are spin-degenerate and disperse similarly to quantum well states \cite{Tchoumakov2017}. A comparison of $E-k_{||}$ cuts slightly away from the \bm{$\Gamma$} point ($k_y=0.073$ \AA$^{-1}$) for the 50 BL and 400 BL films are given in Section S5 of \cite{Supp} along with a \bm{$\Gamma_{002}$} measurement of a 50 BL film to further show the correspondence between the bands in 13 BL, 50 BL, and 400 BL films. 

Measurements of P$_\textrm{Y}$ are shown in Figs. 8(b,c) where a similar spin polarization is seen as for the 400 BL films. The lineshape of the spin polarization in SS1 is much more asymmetric, possibly from the reduced linewidth of SS1 in this film as compared to the 400 BL films. The peak of the SS2 polarization is slightly below E$_\textrm{F}$, compared to the same $k_x$ in 400 BL films where it is at or above E$_\textrm{F}$. This is due to the difference in chemical potentials between the two samples (the thinner film is more electron-doped). As with the 400 BL films, there is no spin polarization in the $\hat{x}$ and $\hat{z}$ directions. The quantum well states do not appear to be spin-polarized, as expected. SS1 and SS2 retain their ideal orthogonal spin-momentum locking and have a clockwise and counterclockwise, respectively, helicity as is the case in 400 BL films. Quantum confinement then does not strongly change the spin texture of SS1 and SS2, but does induce quantum well states and appears to modify the self-energy of SS1 as well. 

\subsection{Potential origins for the surface states observed}\label{ssec:origins}
We observe the presence of two distinct surface states with opposite spin-momentum locked helical spin textures in $\alpha$-Sn, independent of film thickness. We now discuss both the topologically trivial and the non-trivial potential origins of these surface states. Various surface states have been proposed to exist in this system including Dyakonov-Khaetskii states \cite{Dyakonov1981,Khaetskii2022}, Volkov-Pankratov states \cite{Tchoumakov2017,Volkov1985}, states derived from the surface reconstruction of $\alpha$-Sn(001) \cite{Lu1998}, the topological surface state \cite{Ohtsubo2013,Barfuss2013}, and the coexistence of a topological surface state with hybridized Rashba-split surface states \cite{Chen2022}. Zeroth order Volkov-Pankratov states are equivalent to the Dirac-like topological surface state. Higher order Volkov-Pankratov states are spin degenerate and outside of the bulk continuum, and thus cannot be the origin of the spin-polarized SS2. Surface states derived from the reconstruction of $\alpha$-Sn are spin degenerate, and thus cannot be attributed to SS1 or SS2. The quantum well states visible in the 50 BL film could potentially be attributed to Volkov-Pankratov states; The measured states are in the bulk continuum and thus surface resonances rather than surface states, disagreeing directly with the Volkov-Pankratov predictions \cite{Tchoumakov2017}. As already discussed, these quantum well states also share band features with the heavy hole band they are derived from.

Dyakonov-Khaetskii states are spin-polarized, as we observe in SS1 and SS2, however they too exist outside of the bulk continuum rather than inside \cite{Khaetskii2022,Dyakonov1981}. These states modify the dispersion of the linear spin-polarized topological surface state and are sensitive to coupling with the heavy hole band in $\alpha$-Sn, epitaxial strain, band offsets, and film thickness \cite{Khaetskii2022,khaetskii2023}. They connect directly from the bulk heavy hole band ($\Gamma_{8,v}^+$) to the bulk conduction band whereas we observe direct connection of SS1 and SS2 to the inverted conduction band ($\Gamma_{7}^-$) (Fig. 1). However, because these Dyakonov-Khaetskii states are sensitive to so many material parameters, tuning these parameters in the calculation may then result in a picture consistent with the measurements presented here \cite{khaetskii2023}. 

The proposed origin most consistent with the dispersion and spin-texture observed here is the presence of electron-like and hole-like Rashba-split surface states which hybridize to form a Dirac-like topological surface state \cite{Chen2022}. This picture is also consistent with the lack of an upper branch to SS1 in Fig. 2(c) as the outer branches of the Rashba states would form the Dirac-like surface state (SS2), while the inner states would form a lower spin resonance (SS1). However in Chen \textit{et al.} \cite{Chen2022}, the Rashba-split states are calculated/measured to coexist with a topological surface state, which should result in three spin-polarized surface states rather than the two spin-polarized surface states observed here. The helicity of spin-momentum locking in SS1 and SS2 is also inverted from that expected in the minimal model in Chen \textit{et al.} \cite{Chen2022}. In a slightly different mechanism of Rashba-split surface state hybridization, a \textit{single} pair of Rashba-split bands hybridize and result in only two sets of spin-polarized bands (one spin resonance and one Dirac-like surface state), rather than three.  On the other hand, electron- or hole-like Rashba-split surface states could hybridize \textit{with} a pre-existing topological surface to form the observed spin polarization as well (similar to the case of MnBi$_2$Te$_4$\cite{Lee2023}). 

The measured spin polarization from the initial state of both SS1 and SS2 rules out that either band could be a trivial surface state from the surface reconstruction of the film. However if the Rashba effect is significant,  it is possible that the Rashba splitting is acting on such a trivial surface state. The dispersion of surface bands derived from the surface reconstruction of $\alpha$-Sn(001) has been calculated under a number of bonding and domain configurations, but these bands generally switch from a surface resonance at low $k$ to a surface state at high $k$ \cite{Lu1998}. This effect is not observed in the various $E-k$ slices or constant energy contours presented in this work for SS1 and SS2. Surface states derived from the surface reconstruction are not calculated to have connection to the inverted conduction band ($\Gamma_{7}^-$), while this direct connection is observed in all of the films studied here. However it has been shown that Rashba splitting can effect the connection between surface and bulk states \cite{Seibel2015}.
To our knowledge no present calculation or minimal model fully matches the dispersion and spin polarization of surface states observed in this work, although some form of a hybridized Rashba surface state picture \cite{Chen2022,Jozwiak2016,Lee2023} is the most consistent. In this picture SS1 is a lower spin resonance, while SS2 is a Dirac-like topologically non-trivial surface state.
\begin{figure}[ht]
\includegraphics{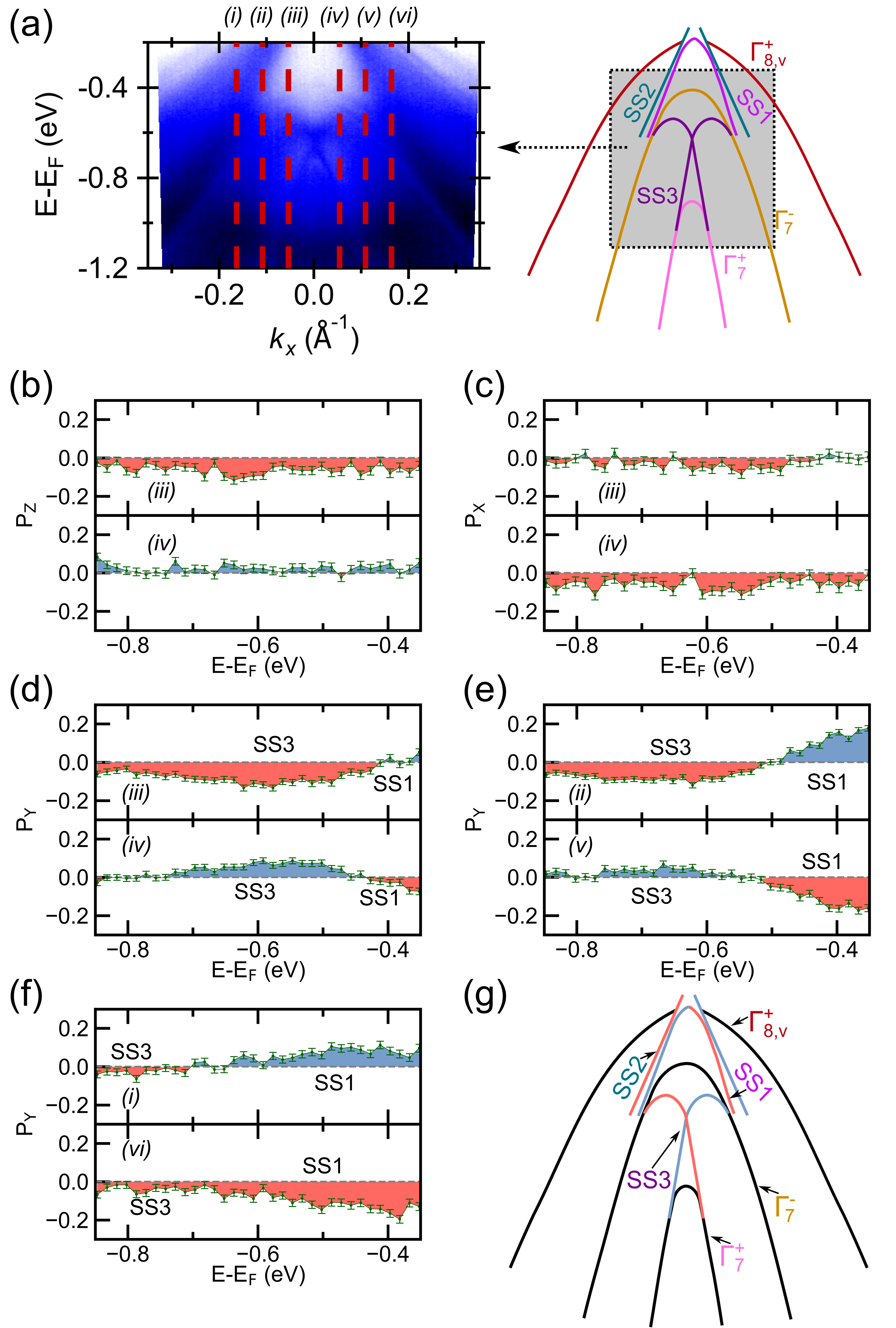}
\caption{\label{Fig9} Spin polarization measurements of SS3. (a) ARPES measurement near SS3 of 400 BL $\alpha$-Sn(001) (400 BL-A) with the EDCs along which spin polarization was measured indicated with the red dashed line. (b) P$_\textrm{Z}$ and (c) P$_\textrm{X}$ measured at $k_{x}=\pm0.054$ \AA$^{-1}$. P$_\textrm{Y}$ measured at (d) $k_{x}=\pm0.054$ \AA$^{-1}$, (e) $k_{x}=\pm0.109$ \AA$^{-1}$, and (f) $k_{x}=\pm0.163$ \AA$^{-1}$. (g) Measured $P_\textrm{Y}$ for surface states 1, 2 and 3. All measurements were made at $h\nu$=21 eV with $p$-polarization. Raw SARPES data is given in Section S3 of \cite{Supp}.}
\end{figure}
\subsection{Spin polarization in the topological surface state from double band inversion in $\alpha$-Sn}\label{ssec:double}

We previously discussed disagreement over the exact shape of the inverted conduction band ($\Gamma_{7}^-$) in Section \ref{ssec:F12}, leading to uncertainty as to the true nature of SS3.  While our spin-integrated ARPES measurements agree with the proposed model in Ref.  \cite{Rogalev2017}, here we seek to measure the spin polarization of this state to further confirm the secondary band inversion and show that SS3 is indeed topologically non-trivial. The EDCs along which spin polarization was measured are shown in Fig. 9(a), where the coexistence of \bm{$\Gamma$} point-like bulk bands and SS3 is clear (Section \ref{ssec:fitting}). The innermost cuts cross the upper branch of the surface state, just touching the lower branch. The middle cuts cross only the upper branch of SS3, while the outermost cuts should mostly interact with the $\Gamma_{7}^-$ band (the cuts are past the $k_x$ extent of SS3 estimated in Fig. 1 and Ref.  \cite{Rogalev2017}). At the innermost cuts (Fig. 9(b,c)) there is no meaningful spin polarization in the $\hat{x}$ or $\hat{z}$ directions.  In Fig. 9(d), a non-zero spin polarization is clearly evident in P$_\textrm{Y}$, centered 600 meV below E$_\textrm{F}$. This spin polarization then has the ideal orthogonal spin-momentum locking expected of a true topological surface state. The SS1 spin polarization is visible at binding energies lower than 400 meV. The upper branch of SS3 has an inverted polarization compared to the lower branch of SS3, implying the lower branch of SS3 has the same spin helicity as SS1 (again inverted from that typically associated with a Dirac-like surface state). In SS1, this helicity (inverted from that expected in a Dirac cone) was shown to not be a function of polarization of light in contrast to what is seen in Bi$_2$Se$_3$ where the polarization controls the spin texture \cite{Jozwiak2011,Jozwiak2013}. However, we have not performed SARPES measurements of the spin texture of SS3 using other polarizations of light or unpolarized light and thus cannot make this same attribution.

The spin polarization decreases slightly farther from the node of SS3 (Fig. 9(e)) and almost fully vanishes when the cut is centered far away from the node of SS3 (Fig. 9(f)).  The measured spin polarization in Fig. 9(f) is dominated by SS1. The spin-polarized SS3 then has some finite $k$ extent -- close to that indicated by the outermost constant momentum slices in Fig. 9(a) -- whereupon it joins the $\Gamma_{7}^-$ band. The \textbf{M}-shaped SS3 is truly a spin-polarized topological surface state with the expected spin-momentum locked spin texture, in conjunction with the spin-polarized surface states SS1 and SS2 (Fig. 9(g)).

\section{Conclusion}
Through detailed spin- and angle-resolved photoemission spectroscopy measurements, we have found several essential clarifications to both the bulk and surface electronic structure of compressively strained $\alpha$-Sn/InSb(001). By excluding the use of extrinsic surface and/or bulk dopants, we isolated the behaviors observed to be intrinsic to $\alpha$-Sn. We have confirmed the presence of a spin-polarized surface state deep below the valence band maximum and found that there is no significant warping of the inverted conduction band, in contrast to many calculations. We have also observed the presence of only two surface states near the valence band maximum across a range of film thicknesses, both of which have their crossing points above the valence band maximum in ultrathin films; the third state sometimes seen in low photon energy ARPES measurements, for our films, was consistent with the heavy hole dispersion at the \bm{$\Gamma$} point for both 13 and 400 BL $\alpha$-Sn films. Most importantly, both of these near-VBM surface states were observed to be spin-polarized with the ideal orthogonal spin-momentum locking but opposite helicities. We find that the inner spin-polarized surface state is likely a lower spin resonance (from a form of hybridization with Rashba states), while the outer spin-polarized surface state is the topologically non-trivial Dirac-like surface state.

Our results exemplify the complexity of not only the electronic structure of $\alpha$-Sn, but also the measured photoemission spectra. Few calculations predict the dispersion of the inverted conduction band or the dispersion and spin textures of the surface states observed in this work. A better agreement between theory and experiment would help with the understanding of $\alpha$-Sn such that a more deterministic control of the topological phase is possible. As is, the clarification to the electronic structure of $\alpha$-Sn reported here sheds light on the results of other measurement techniques which are not sensitve to the full band dispersion. Furthermore, the existence of oppositely spin-polarized surface states terminating above the valence band maximum could allow gate or dopant controlled tuning of the chemical potential in $\alpha$-Sn to increase the already remarkable spin-charge conversion efficiency in this system, while also minimizing the contributions of the parasitic bulk channel.

\begin{acknowledgments}

The growth and later ARPES studies were supported by the Army Research Laboratory (W911NF-21-2-0140 and W911NF-23-2-0031). The initial vacuum suitcase construction and initial ARPES measurements were supported by the US Department of Energy (DE-SC0014388). The UC Santa Barbara NSF Quantum Foundry funded via the Q-AMASE-i program under award DMR-1906325 support was used for further development of the vacuum suitcases. This research used resources of the Advanced Light Source, which is a DOE Office of Science User Facility under contract no. DE-AC02-05CH11231.  Use of the Stanford Synchrotron Radiation Lightsource, SLAC National Accelerator Laboratory, is supported by the U.S. DOE, Office of Science, Office of Basic Energy Sciences under Contract No. DE-AC02-76SF00515. The research reported here made use of the shared facilities
of the Materials Research Science and Engineering Center (MRSEC) at UC Santa Barbara: NSF DMR–2308708. The authors would like to further thank A. Khaetskii, G. J. de Coster, J. W. Harter, P. J. Taylor, A. M. Kiefer, P. A. Folkes, and O. A. Vail for fruitful discussions.
\end{acknowledgments}


\bibliography{export}


\end{document}